\documentclass[twocolumn]{bmcart}

\usepackage{amsthm,amsmath}
\RequirePackage[numbers]{natbib}
\RequirePackage{hyperref}
\RequirePackage{scalerel}
\RequirePackage{tikz}
\usepackage[utf8]{inputenc} 

\startlocaldefs
\endlocaldefs

\begin{document}

\begin{frontmatter}

\begin{fmbox}
\dochead{Research}


\title{Efficient emotion recognition using hyperdimensional computing with combinatorial channel encoding and cellular automata}


\author[
  addressref={aff1},                   
  corref={aff1},                       
  email={allymenon@berkeley.edu}   
]{\inits{A.}\fnm{Alisha} \snm{Menon}}
\author[
  addressref={aff1},
  email={anirudhn@berkeley.edu}
]{\inits{A.}\fnm{Anirudh} \snm{Natarajan}}
\author[
  addressref={aff1},
  email={revaagashe23@berkeley.edu}
]{\inits{R.}\fnm{Reva} \snm{Agashe}}
\author[
  addressref={aff1},
  email={daniel.s@berkeley.edu}
]{\inits{D.}\fnm{Daniel} \snm{Sun}}
\author[
  addressref={aff1},
  email={maristio@berkeley.edu}
]{\inits{M.}\fnm{Melvin} \snm{Aristio}}
\author[
  addressref={aff1},
  email={harrisonliew@berkeley.edu}
]{\inits{H.}\fnm{Harrison} \snm{Liew}}
\author[
  addressref={aff1},
  email={ysshao@berkeley.edu}
]{\inits{Y.S.}\fnm{Yakun Sophia} \snm{Shao}}
\author[
  addressref={aff1},
  corref={aff1},  
  email={jan\_rabaey@berkeley.edu}
]{\inits{J.M.}\fnm{Jan M.} \snm{Rabaey}}


\address[id=aff1]{
  \orgdiv{Department of Electrical Engineering and Computer Science},             
  \orgname{University of California at Berkeley},          
  \city{Berkeley, CA},                              
  \cny{USA}                                    
}






\begin{abstractbox}

\begin{abstract} 
In this paper, a hardware-optimized approach to emotion recognition based on the efficient brain-inspired hyperdimensional computing (HDC) paradigm is proposed. Emotion recognition provides valuable information for human-computer interactions, however the large number of input channels ($>$200) and modalities ($>$3) involved in emotion recognition are significantly expensive from a memory perspective. To address this, methods for memory reduction and optimization are proposed, including a novel approach that takes advantage of the combinatorial nature of the encoding process, and an elementary cellular automaton. HDC with early sensor fusion is implemented alongside the proposed techniques achieving two-class multi-modal classification accuracies of $>$76\% for valence and $>$73\% for arousal on the multi-modal AMIGOS and DEAP datasets, almost always better than state of the art. The required vector storage is seamlessly reduced by 98\% and the frequency of vector requests by at least 1/5. The results demonstrate the potential of efficient hyperdimensional computing for low-power, multi-channeled emotion recognition tasks.
\end{abstract}




\end{abstractbox}
\end{fmbox}

\end{frontmatter}



\section{Introduction}
Affective computing for informed human-computer interaction (HCI) is an area of growing research interest \cite{yin2017recognition}. Traditional interfaces such as keyboards and mouse are limited to conveying explicit information; the HCI experience can be enhanced through the inclusion and interpretation of additional implicit information \cite{zeng2008survey}. For example, context-dependent human behavioral patterns can be learned and used to inform feedback systems of user intention in a wide variety of applications such as driver warning systems, smart environments, automated tutoring systems, etc. \cite{zeng2008survey},\cite{shojaeilangari2015robust},\cite{kapoor2007automatic}. Providing computers with emotional skills will allow them to intelligently react to subtle user context changes such as emotional state \cite{picard2001toward}. 

Though a common approach is interpreting audio-visual signals such as facial expressions and voices, these may not be the primary source of expression. Emotion is not always easily observable, rather it requires inference through various behavioral observations and physiological indices which together can provide sufficient information \cite{cohn2007foundations}. Many existing datasets collected for affective computing include various forms of physiological signals to create a comprehensive observation of emotional state \cite{correa2018amigos},\cite{koelstra2011deap}. In the era of Internet-of-things (IoT), advances in wearable devices make the inclusion of various sensing modalities in intelligent HCI applications increasingly feasible \cite{chang2019hyperdimensional}. 

A representation of emotion used for affective computing is the arousal-valence plane \cite{posner2005circumplex}. This model describes discrete emotional states on this space in terms of varying levels of arousal or emotional intensity, and of valence or polarity of emotion (positive or negative). For emotion recognition tasks, these are reduced to high and low arousal and valence values which can, in combination, be used to define the nature of the emotion. 

The emotion recognition system must also be able to address the challenge of multimodal classification which results from the inclusion of diverse physiological sensors \cite{wang2018entropy}. For this work, the AMIGOS and DEAP datasets were selected specifically for the large number of sensor channels and modalities. The AMIGOS dataset contains electroencephalogram (EEG), galvanic skin response (GSR) and electrocardiogram (ECG) sensors \cite{correa2018amigos}. The DEAP dataset includes EEG, Electrooculography (EOG), Electromyography (EMG), GSR, blood volume pressure (BVP), temperature and respiration sensors \cite{koelstra2011deap}.

Previous work on multi-modal fusion for the AMIGOS dataset includes Fisher’s linear discriminant with Gaussian Naive Bayes, which was shown to achieve F1 scores of 57\% and 58.5\% on valence and arousal \cite{correa2018amigos}, \cite{chang2019hyperdimensional}. Wang et al. implemented recursive feature elimination (RFE) with a support vector machine (SVM) and obtained 68\% and 66.3\% accuracy on valence and arousal \cite{wang2018entropy}. Wang et al. also implemented Extreme Gradient Boosting (XGBoost) for accuracies of 80.1\% and 68.4\% on valence and arousal. Siddharth et al. used extreme learning machines (ELM) for accuracies of 83.9\% and 82.7\% on valence and arousal \cite{siddharth2019utilizing}. Previous binary classification multimodal fusion approaches for the DEAP dataset include a restricted boltzmann machine (RBM) with an SVM classifier, with accuracies of 60.7\% and 64.6\% for valence and arousal, respectively \cite{shu2017emotion}. Wang et al. used a deep belief network (DBN) through multi-layer RBMs for valence and arousal accuracies of 51.2\% and 68.4\% \cite{wang2013modeling}. Yin et al. used a multiple-fusion-layer based ensemble classifier of stacked autoencoder (MESAE) for accuracies of 76.2\% and 77.2\% for valence and arousal \cite{yin2017recognition}.

Since emotion recognition can provide valuable information for HCI, a hardware-efficient platform that allows for extended-use, on-board classification, would increase the feasibility of long-term wearable monitoring and thus increase the scope of potential feedback applications. While previous work shows strong results for the AMIGOS and DEAP datasets in terms of classification accuracy, the ease of hardware implementation for training and inference are not considered while designing the models; these methods have high computational complexity that reduces implementation feasibility on resource-limited wearable platforms. SVMs, for example, while demonstrating high accuracy, are challenging to implement efficiently on hardware, and demonstrate a trade-off between precise accuracy and meeting hardware constraints \cite{afifi2020fpga}, \cite{montagna2018pulp}. In addition, multimodal fusion approaches require parallel encoding schemes prior to the fusion point which further increase the complexity creating a bottleneck for real-time wearable classification.   

To address this, in this work Brain-inspired Hyperdimensional Computing (HDC) is used for emotion recognition. HDC is an area of active research that has been successfully demonstrated for classifying physiological signals such as the wearable EMG classification system implemented from Moin et al. that achieves 97.12\% accuracy in recognizing 13 different hand gestures \cite{moin2020wearable}, the iEEG seizure detection system developed by Burrello et al. \cite{burrello2019laelaps}, and the EEG error-related potentials classification for brain-computer interfaces implemented by Rahimi et al. \cite{rahimi2017hyperdimensional}. It is based on the idea that human brains do not perform inference tasks using scalar arithmetic, but rather manipulate large patterns of neural activity. These patterns of information are encoded in binary hypervectors, with dimensions ranging in the thousands to ensure that any two random HVs are likely to be nearly orthogonal to each other \cite{kanerva2009hyperdimensional}. There are three operations that are performed on these hypervectors: bundling, binding, and permutation. Bundling is a component-wise add operation across input vectors, binding is a component-wise XOR operation, and permutation is a 1-bit cyclical shift. The simplicity of these operations suggests that HDC is very hardware efficient, as confirmed in previous work \cite{datta2019programmable}, \cite{montagna2018pulp}. Montagna et al. demonstrated that HDC computing achieved 2x faster execution and lower power at iso-accuracy on an ARM Cortex M4 compared to an optimized SVM \cite{montagna2018pulp}. 
\begin{figure*}
  \centering
  \includegraphics[scale=0.265]{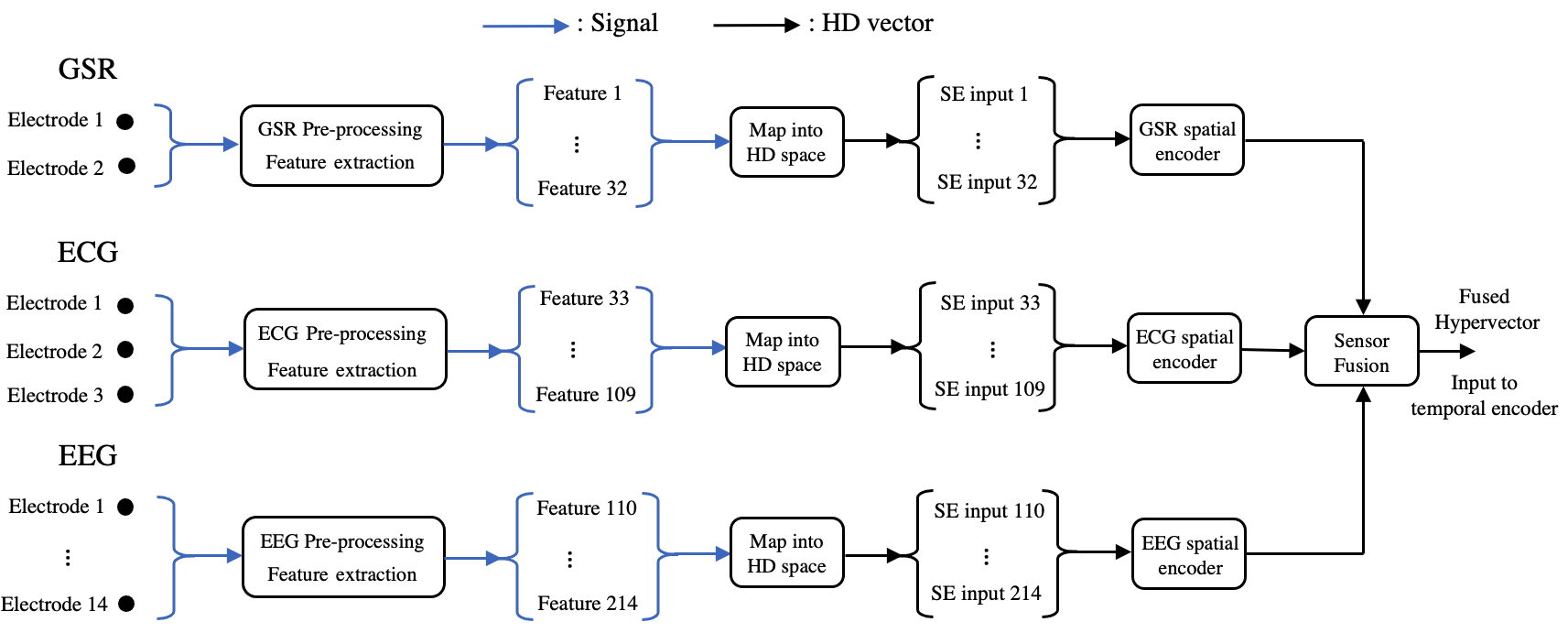}
  \caption{The sensor fusion datapath from electrodes to a fused hypervector for the three-modality emotion recognition system used in AMIGOS with GSR, ECG, and EEG sensor inputs. The sensor inputs are pre-processed into a set of features which are then mapped into the HD space to create a set of spatial encoder (SE) inputs. These vectors are encoded within each modality, and then finally fused together to create one vector representing information from all of the channels. The process is detailed in Section 2}
  \label{big picture}
\end{figure*}

HDC has additional properties that demonstrate its potential for a wearable emotion recognition system. Previous work by Chang et al. developed a baseline, unoptimized architecture for emotion classification on the AMIGOS dataset, which was able to achieve valence and arousal accuracies of 83.2\% and 70.1\%, respectively, demonstrating higher performance than SVM, XGB and gaussian naive bayes for all amounts of training data \cite{chang2019hyperdimensional}. HDC encodes information in the same form no matter the type, number or complexity of the inputs. This is accomplished through basic vectors (items) that are random, and typically stored in an item memory (a codebook). Each channel is assigned a unique item memory vector, and feature values are typically encoded through a discretized mapping to additional unique hypervectors representing values within a set range. Each stream of information is encoded into this representation as shown in Figure \ref{big picture}, which lends HDC well to sensor fusion.  

HDC inherently binds features extracted from various physiological data streams. This suggests early fusion with reduction of parallel encoding schemes will have little effect on its accuracy, breaking the complexity-accuracy tradeoff. HDC offers a reduction of computation and memory requirements in contrast to traditional machine learning models, demonstrated by Montagna et al. \cite{montagna2018pulp}. It also offers the ability to use the same hardware for training \& inference, rapid training, and robustness to noise/variations in input data making it a viable choice for wearable, hardware-constrained sensor fusion applications. 

Datta et al. synthesized an implementation of a generic HDC application-specific integrated circuit (ASIC) processor that provided a power breakdown between the various blocks involved \cite{datta2019programmable}. The item memory, which stores channel identification vectors, contributed the most, 42\%, to the overall power of the processor. For the emotion recognition-specific application, the large number of channels ($>$200) and modalities ($>$3) requires advance storage of a correspondingly large number of unique vectors in the item memory. More channels translates into more memory. This would result in memory storage consuming $\sim$50\% of the overall processor power similar to \cite{datta2019programmable}. Reduction of memory usage would allow HDC to meet stricter power/complexity constraints, improving its potential for implementation on wearable platforms.

In this work, use of pseudo-random vector generation through computation using a cellular automata is proposed and implemented for this purpose. A cellular automata consists of a grid of cells which evolve with complex, random behavior over time through a set of discrete computations using the current state and that of nearby cells \cite{kleyko2017no}. Cellular automata rule 90 assigns the next state in a method shown to be equivalent to an XOR of the two nearest cells \cite{kleyko2017no}. 
For a hypervector, each cell represents a single bit and rule 90 can be implemented through XOR of the cyclical left-shift and cyclical right-shift of the original vector. If HV$_\textrm{n}$ is the hypervector state at step n, and $\rho$ is the cyclical shift operation (+1 for right, -1 for left), then HV$_{\textrm{n}+1}$ can be generated by:
\begin{equation}
\textrm{HV$_{\textrm{n}+1}$} = \rho^{+1}(\textrm{HV$_\textrm{n}$}) \oplus \rho^{-1}(\textrm{HV$_\textrm{n}$})
\end{equation}
These operations are vectorizable and computationally minimal. The emotion recognition architecture uses a fixed sequential channel (item) access pattern, therefore, this technique, with which the item memory vectors are sequentially-evolving, can be used. Cellular automata grid sizes over 24 have been shown to generate new degrees of freedom for more than 10$^3$ steps before saturating \cite{kleyko2020cellular}. Hypervectors, with tens of thousands of cells in the grid, provide linearly longer randomization periods; this is sufficient for most applications including emotion recognition. 
Using a single random seed vector, full-dimension random item memory hypervectors can be generated during the encoding process instead of being precomputed and stored. With this approach, the memory is constant regardless of the number of channels. 

In this work, an HDC architecture is designed specifically targeting sensor fusion using an early fusion approach. This reduces the parallel encoding paths previously used for HDC sensor fusion to a single one by taking advantage of HDC's inherent projection of features into large-capacity hypervector representations. Methods for memory reduction and optimization are proposed and implemented, including a novel approach that takes advantage of the combinatorial nature of the HDC encoding process, and the usage of an elementary cellular automata with rule 90 together to reduce vector storage and request frequency. Finally, hypervector dimension reduction is further explored as a method of comprehensive reduction. Results are reported for both the DEAP and AMIGOS datasets.
\begin{figure*}
  \centering
  \includegraphics[scale=0.33]{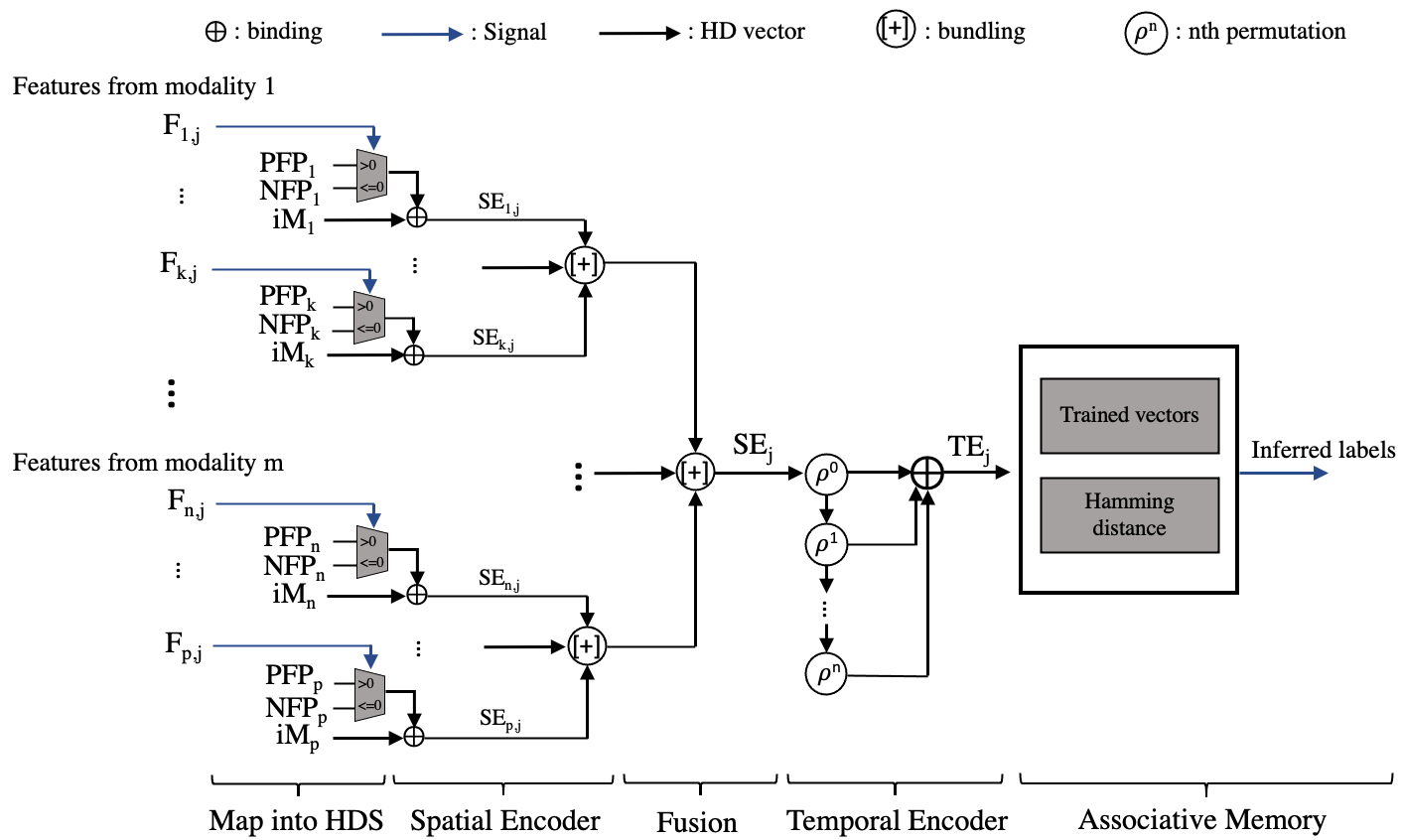}
  \caption{HDC early fusion detailed architecture for m modalities with the four main blocks: map into HDS, spatial encoder, temporal encoder for n-gram of size n+1, and associative memory. Sensor fusion occurs early in the datapath, directly after the spatial encoder. }
  \label{early fusion}
\end{figure*}

\section{Methods}
\subsection{HDC early fusion architecture}
The HDC physiological architecture includes four main blocks: the map into the hyperdimensional space (HDS), the spatial encoder, the temporal encoder, and the associative memory as shown in Figure \ref{early fusion}. The first block maps incoming data into HDS using an item memory or a generator. HDC depends on the pseudo-orthogonality of random vectors to be able to distinguish between various classes; a random vector will be nearly orthogonal to another random vector in the hyperdimensional space. Random vectors are used for the channel item memory vectors so that the source channel of a feature value can be included as information in the encoding process. These are stored in an item memory (iM).

In order to encode feature values, in this implementation, additional feature projection vectors are randomly generated for each channel and stored as well. In traditional architectures, the feature projection vector \{-1, 0, 1\} is multiplied by the feature value and then binarized by reducing the positive values in the vector to 1s, and the zeros and negative values to 0s. This process can be simplified to multiplexers selecting between a pre-generated random negative or positive binary feature projection vector depending on the feature value’s sign to eliminate computationally-expensive multipliers. This allows the feature projection vectors to maintain pseudo-orthogonality but have the same sparsity as the item memory vectors, making them interchangeable. As a result, the feature projection vectors can also be stored in the item memory instead of separately.

In the spatial encoder, the binding operation (XOR) is utilized to generate a spatially encoded hypervector for each channel. If iM$_i$ represents the item memory vector for channel $i$ and FP$_{i,j}$ represents the feature projection vector selected for channel $i$ for sample $j$, then the spatially encoded hypervector for sample $j$ SE$_{i,j}$ is computed as:
\begin{equation}
\textrm{SE$_{i,j}$} = \textrm{iM$_{i}$} \oplus \textrm{FP$_{i,j}$}
\end{equation}

To develop a complete hypervector, the bundling operation (vertical majority count across vectors) combines the many spatially encoded hypervectors within a sensor modality. If the sensor modality $m$ has $k$ channels and the bundling operation is represented as $+$, SE$_{m,j}$ is computed as:  
\begin{equation}
\textrm{SE$_{m,j}$} = \textrm{(iM$_{1}$} \oplus \textrm{FP$_{1,j}$)} + ... + \textrm{(iM$_{k}$} \oplus \textrm{FP$_{k,j}$)}
\end{equation}

\begin{figure}
  \centering
  \includegraphics[scale=0.35]{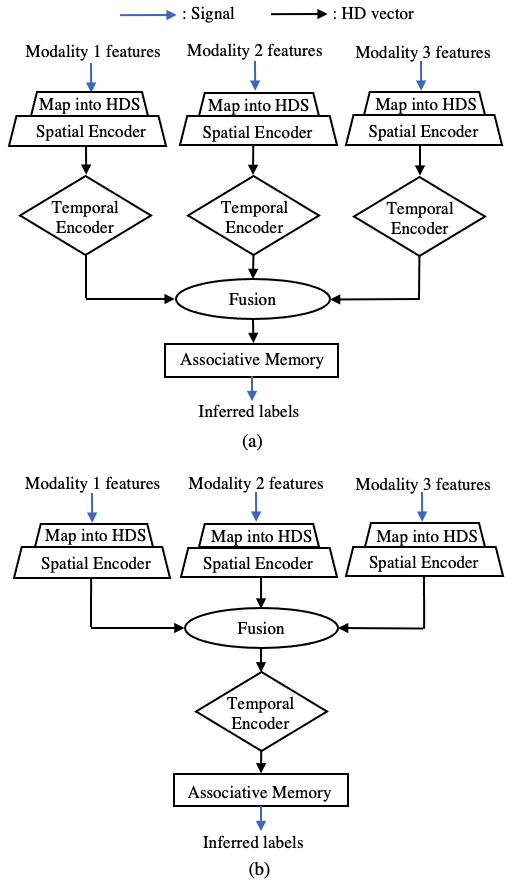}
  \caption{HDC (a) early fusion and (b) late fusion architectures for a three-modality emotion recognition system. The late fusion architecture fuses after the temporal encoder, resulting in 3 parallel temporal encoders - one per modality. In comparison, the early fusion architecture fuses before the temporal encoder, resulting in only 1 temporal encoder.}
  \label{earlyvlate}
\end{figure}

Because emotion recognition involves various sensor modalities, it requires fusion. Previous sensor fusion implementations fused after the temporal encoder, but in this work, an early fusion approach is taken, which fuses the modalities directly after the spatial encoding process. Therefore, this architecture requires only a single temporal encoder as opposed to one per modality, as shown in Figure \ref{earlyvlate}. This reduces the parallel encoding paths while still allowing each modality to be weighted equally instead of by number of features. If there are $m$ sensor modalities, the fused spatially encoded hypervector for sample $j$ is:
\begin{equation}
\textrm{SE$_{j}$} = \textrm{SE$_{1,j}$} + \textrm{SE$_{2,j}$} + ... + \textrm{SE$_{m,j}$}
\end{equation}

HDC also has the ability to encode temporal changes through the use of n-grams based on a sequence of $N$ samples. This is invaluable for physiological signals that are time-varying as it allows for the capturing of time-dependent emotional fluctuations within the same class or between segments of the same class. The permutation operation (cyclical shift, represented as $\rho$) is used to keep track of previous samples. Hypervectors coming from the spatial encoder are permuted and then bound with the next hypervector $N$ times in the temporal encoder. This results in an output that observes changes over time, TE${_j}$, that can be computed as: 
\begin{equation}
\textrm{TE$_{j}$} = \textrm{SE$_{j}$} \oplus \rho^{+1}\textrm{(SE$_{j-1}$)} \oplus ... \oplus \rho^{+(N-1)}\textrm{(SE$_{j-(N-1)}$)}
\end{equation}

During the training process, many such encoded hypervectors are generated, bundled to represent a class and then stored into the final block, the associative memory. During inference, the encoded hypervector is compared against each trained hypervector using Hamming distance. For binary vectors, this involves an XOR and then popcount. The comparison with least distance is the inferred label.
\begin{figure*}
  \centering
  \includegraphics[scale=0.35]{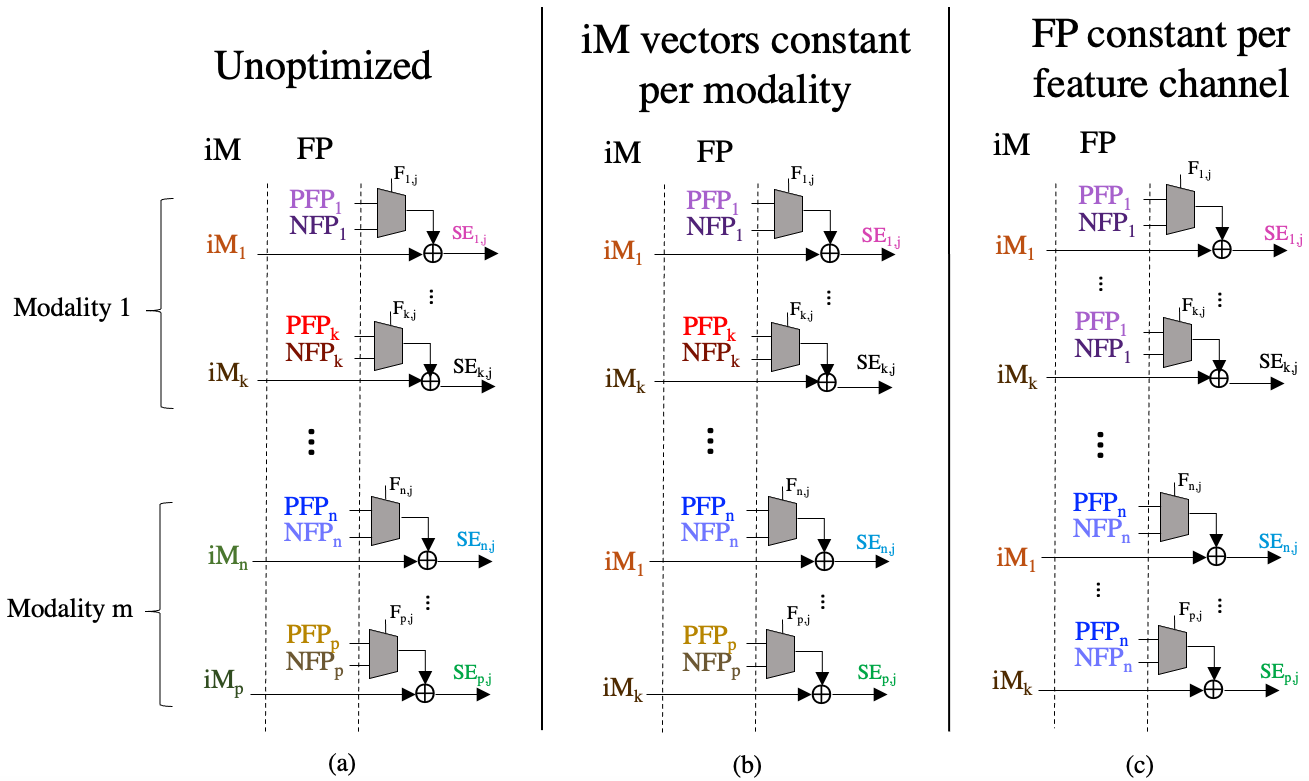}
  \caption{iM and FP vectors used to map into HDS to generate unique SE vectors per channel for (a) `unoptimized' with distinct iM and FP vectors for all channels, (b) `iM vectors constant per modality' with the same iM vectors between different modalities, and (c) `FP constant per feature channel' with the same FP vectors between channels of the same modality.}
  \label{memopt}
\end{figure*}

\subsection{Implementation}
The HDC early fusion architecture is implemented on both the AMIGOS and DEAP datasets with a standard  dimension of 10,000 for the full datapath in the baseline implementation. In the AMIGOS study, GSR, ECG and EEG signals were measured for 33 subjects as they watched 16 videos \cite{correa2018amigos}. Each video for each subject was classified to have either led to a positive or negative emotion (valence), and the strength of the emotion was classified as either strong or weak (arousal). From the 3 modalities, Wang et al. selected 214 time and frequency domain features relevant to accurate emotion classification \cite{wang2018entropy}. GSR has 32 features, ECG has 77 features, and EEG has 105 features. Similar features are used in this work. The data for all 33 subjects was appended and a moving average of 15 seconds over 30 seconds was applied. The signals were scaled to be between $-1$ and $+1$ to meet the HDC encoding process and downsampled by a factor of 8 for more rapid processing and usage of the HDC classification algorithm. Previous work uses the leave-one-subject-out approach to evaluate performance, this was also implemented for the early fusion architecture \cite{correa2018amigos}, \cite{wang2018entropy}, \cite{siddharth2019utilizing}. The temporal encoder was tuned and an optimal n-gram of 3 feature windows was selected. For both datasets, transitionary ngrams (those with samples from both classes) were excluded from training and testing. 

The DEAP study was collected in a similar format as the AMIGOS with 32 subjects watching 40 one-minute highlight excerpts of music videos selected to trigger distinct emotional reactions, however it contains more extensive sensor modalities: EEG, EMG, GSR, BVP, EOG, temperature and respiration \cite{koelstra2011deap}. The arousal and valence scores were self-assessed by the participants on a scale between 1-9. A binary classification system is maintained for high and low valence and arousal by thresholding the scale at 5. Preprocessing and feature selection were done using the TEAP toolbox which selected time and frequency domain features for 5 of the modalities based on previous work in those areas \cite{soleymani2017toolbox}. EMG has 10 features, EEG has 192 features, GSR has 7 features, BVP has 17 features, and respiration has 12 features per video. This results in 40 samples with a total of 238 features from 5 modalities per subject. The signals were then scaled to be between $-1$ and $+1$ for the HDC encoding scheme. Previous work for this dataset does training and inference independently by subject which was adopted in this work as well \cite{shu2017emotion}, \cite{koelstra2011deap}, \cite{yin2017recognition}. Typically, 90\% of the dataset is used for training per subject with the remaining 4 videos used for testing. For HDC, due to the inclusion of the temporal encoder, this would result in limited number of inferences leading to imprecise classification accuracies. As a result, the size of the training set was decreased to be 80\% of the dataset with 20\% used for testing. A temporal n-gram of 3 was selected for this dataset as well. 

\subsection{Memory optimization}
For both the AMIGOS and DEAP datasets, there are over 200 features that need to be spatially encoded. This requires advance storage of 214/238 iM vectors and 420/476 feature projection (FP) vectors - positive (PFP) and negative (NFP) - totalling to 642/714 vectors that need to be stored in the item memory. Use of a unique iM and FP vector set per channel is shown in first column of Figure \ref{memopt}. Without significant reduction of the memory requirements, optimizations of other blocks will provide limited benefits to the overall efficiency.

In the spatial encoder, the iM vector and the FP vector are bound together to form a unique representation containing feature information that is specific to a feature channel. However, both the iM and FP vectors do not need to be unique to the feature channel in order to generate a unique combination of the two. The binding operation will inherently create a vector different, and pseudo-orthogonal to both of its inputs. Therefore, as long as one of these inputs is different for a specific feature channel, the spatially encoded feature channel vector (represented by the SE vectors in Figure \ref{memopt}) will be unique. Using this idea, a set of optimizations were developed and implemented on the DEAP and AMIGOS datasets:
\begin{figure*}
  \centering
  \includegraphics[scale=0.28]{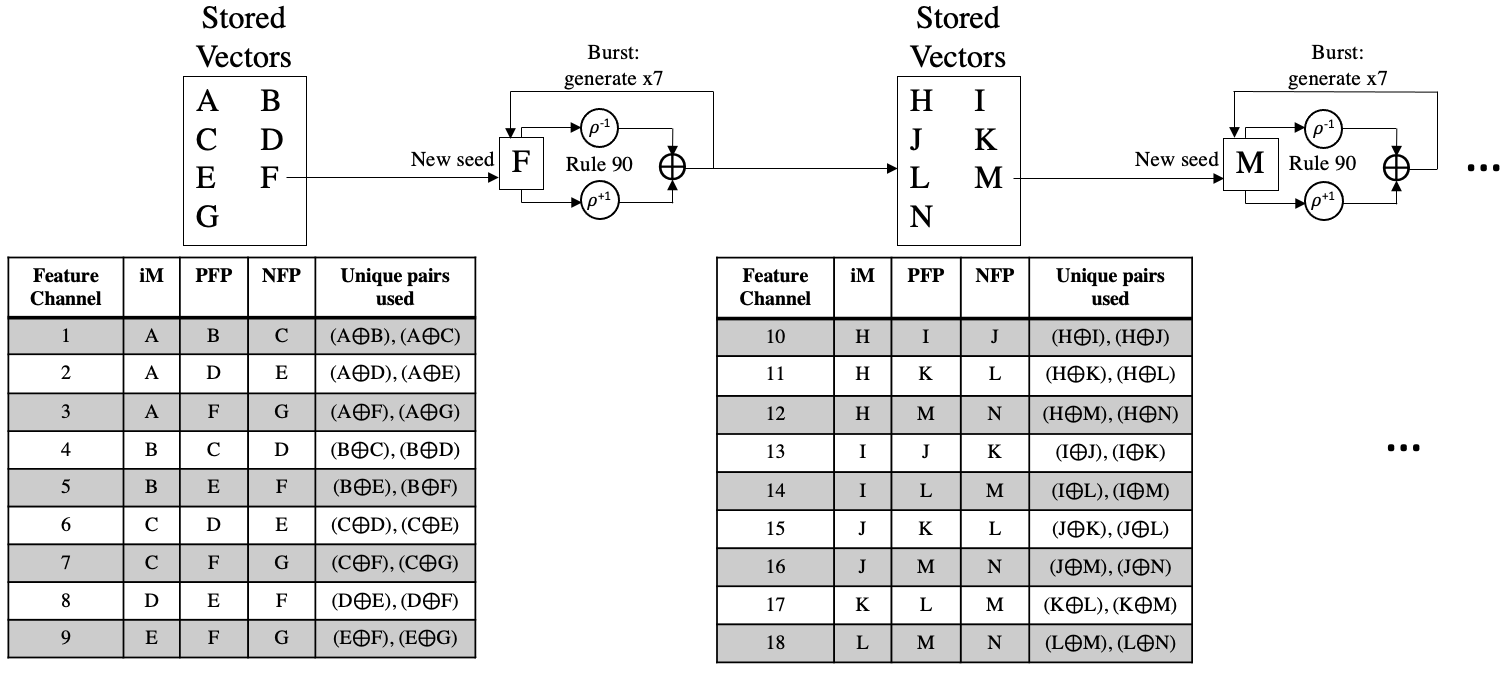}
  \caption{`Combinatorial pairs' feature channel vector set generation demonstrated for 7 stored vectors. iM loops through vector bank after exhausting available sequential pairs for FP. Hybrid method follows by burst re-generating the vector bank with rule 90 so that new combinatorial pairs can be formed for more feature channels. Generation of 18 feature channel vector sets using a bank of only 7 vectors is shown.}
  \label{comb}
\end{figure*}

\textbf{`iM vectors constant per modality'}: the iM is replicated across the various modalities, shown in the second column of Figure \ref{memopt}. If, between each modalities, the FP vectors are different, then orthogonality and input feature channel uniqueness are maintained even if the iM is the same.

\textbf{`FP constant per feature channel'}: though the iM is now the same between each modalities, each feature channel within a modality still has a unique iM vector. Therefore, it is possible to re-use the same FP vectors for every feature channel within a modality, as shown in the third column of Figure \ref{memopt}. This requires maintaining 2 unique FP vectors (PFP and NFP) for each modalities, and unique iM vectors within a modality.

\textbf{`Combinatorial pairs'}: taking this combinatorial binding strategy to its limit, the 2-input binding operation can be used to generate many unique vectors from a smaller set of vectors by following an algorithmic process. Each feature channel requires a distinct set containing an iM vector, and two FP (positive \& negative) vectors: \{iM, PFP, NFP\}. If the vectors for feature channel $1$ are \{A, B, C\}, then the bound pairs that could result from spatial encoding (iM $\oplus$ PFP or iM $\oplus$ NFP) are:
\begin{itemize}
    \item A $\oplus$ B
    \item A $\oplus$ C
\end{itemize}
B $\oplus$ C will not occur because they are both FP vectors. However, it is a unique pairing that could be re-used for another channel. For example, the set for feature channel $2$ could be: \{B, C, D\}. The encoding process would use the following pairings:
\begin{itemize}
    \item B $\oplus$ C
    \item B $\oplus$ D
\end{itemize}
This re-use strategy is the key to saving memory; it can be applied across all channels using a bank of the minimal required vectors, as shown in the first part of Figure \ref{comb}. 
    
Each vector can be paired with every other vector only once in order to maintain orthogonality and paired uniqueness across all feature channel. For a feature channel, one vector (the iM) must have two other available vectors (PFP and NFP) to pair with. With $\left \lfloor{x}\right \rfloor$ defined as the floor function of x, the following equation can be used to calculate the total feature channels, TFC, possible given a bank of $v$ vectors:
\begin{equation}
\label{combinatorial_equation}
    \textrm{TFC} = \sum_{n=1}^{v-2} \left \lfloor{\frac{v - n}{2}}\right \rfloor
\end{equation}
The formula can be derived by looping through each vector in the vector bank and sequentially grouping it with pairs of the other vectors. The generation of feature channel sets can be algorithmic, following the pattern shown in the tables in Figure \ref{comb}.

\textbf{`Rule 90 generation'}: implementation of the cellular automata with rule 90 will allow trading off vector storage with vector generation. If there are $m$ modalities, the first $2\times m$ generated vectors would be used for the PFP and NFP vector for each modality. These would be maintained throughout training and inference resulting in $2\times m + 1$ locally stored vectors including an initial seed vector. However, the rest of the iM vectors would be generated on the fly for each feature channel during the encoding process, requiring no additional vector storage. This is possible because of the fixed access pattern of the iM. The generation process requires use of rule 90 across the entire hypervector, and local storage of the most recently generated vector to use as the next seed. $1$ vector is requested and then generated for each feature channel.  

\textbf{`Hybrid'}: to reduce vector requests and hence the computation for rule90, the last two schemes: `combinatorial paired binding' and `rule 90 generation', can be combined. This hybrid strategy could include burst generation of a small set of vectors which could be locally stored. From this set, combinatorial pairs are assigned to feature channels and spatially encoded. This set can be gradually re-populated with new vectors as the old vectors are exhausted in the encoding process providing new possible pairs. This provides further tradeoff between vector storage and computation. The vector request rate (vector generation requests per feature channel) is minimized when the vector storage is large enough for the combinatorial paired binding scheme alone at which point no generation is required. 

\textbf{`Dimensionality reduction'}: the final method of memory reduction is in the form of hypervector dimension reduction. This changes the size of the entire datapath, impacting both the logic complexity and the memory storage approximately linearly. However, smaller hypervectors also have reduced pseudo-orthogonality - random lower-dimensional vectors are less likely to be nearly orthogonal in the hyperdimensional space than higher-dimensional vectors. The capacity for information is reduced, potentially impacting classification accuracy. This optimization is a tradeoff between algorithm accuracy performance and overall efficiency. The impact of changing dimensions on emotion recognition accuracy for the various memory optimizations is also explored.
\begin{figure*}
  \centering
  \includegraphics[scale=0.37]{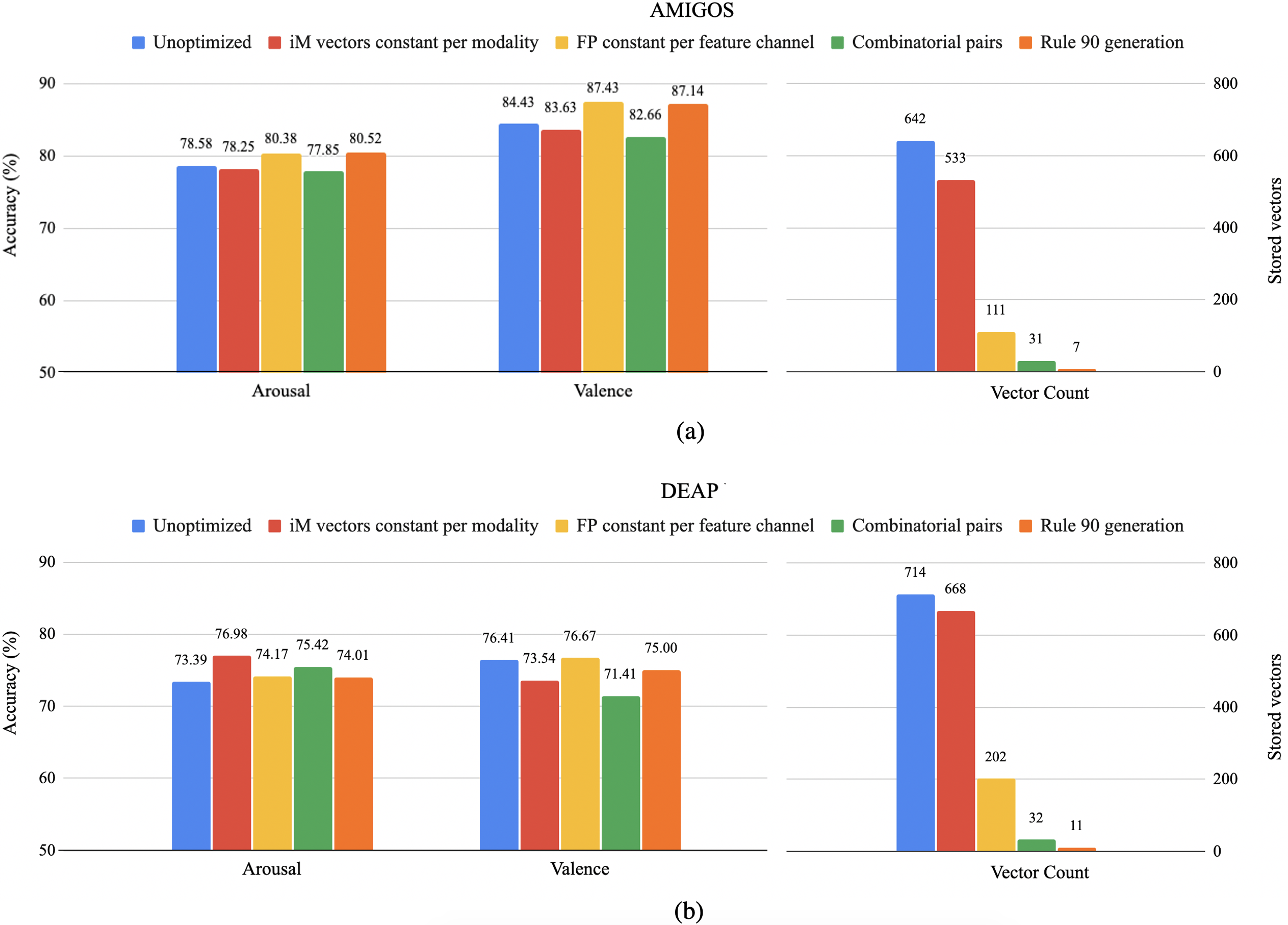}
  \caption{Arousal and valence accuracies and required vector storage for the various memory optimization as compared to unoptimized for (a) AMIGOS and (b) DEAP datasets. Optimizations include `unoptimized' with distinct iM and FP vectors for all channels, `iM vectors constant per modality' with the same iM vectors between different modalities, `FP constant per feature channel' with the same FP vectors between channels of the same modality, and `Rule 90 generation' with generated FP and iM vectors on top of the previous optimizations.}
  \label{memopt_perf}
\end{figure*}

\section{Results}
The HDC early fusion architecture was implemented on the AMIGOS and DEAP datasets for classification of high vs. low arousal and high vs. low valence. HDC early fusion achieved the highest average valence and arousal accuracy on AMIGOS, with the Rule 90 generation encoding method. A comparison against other AMIGOS binary classification multi-modal work using SVM, XGB, Gaussian Naive Bayes (GaussianNB) and ELM is shown in Table \ref{Amigos comparison}. The early fusion encoding process provided a boost of 3.9\% for valence and 10.4\% for arousal from the late fusion HDC architecture previously implemented by Chang et al. \cite{chang2019hyperdimensional} and demonstrates higher average accuracy than state of the art. 

On the DEAP dataset, HDC early fusion achieved the highest average valence and arousal accuracy with the FP constant per feature channel encoding method. A comparison against other DEAP binary classification multi-modal work using GaussianNB, RBM with SVM, MESAE, and DBN is shown in Table \ref{deap comparison}. HDC early fusion accuracy is very comparable with other high performance multi-modal approaches to the DEAP dataset.

\begin{table}[h!]
\caption{AMIGOS classification accuracy comparison table}
\label{Amigos comparison}
  \begin{tabular}{ccc}
    \hline
    \textbf{Method} & \textbf{HV vs. LV (\%)}  &\textbf{HA vs. LA (\%)}\\ \hline
    GaussianNB* \cite{correa2018amigos}\cite{chang2019hyperdimensional} & 57 & 58.5\\
    SVM \cite{wang2018entropy} & 68.0 & 66.3\\
    ELM \cite{siddharth2019utilizing} & 83.9 & 82.8 \\
    XGB \cite{wang2018entropy} & 80.1  & 68.4 \\
    HDC late fusion \cite{chang2019hyperdimensional} & 83.2  & 70.1 \\
    \textbf{HDC early fusion} & 87.1 & 80.5 \\ \hline
  \end{tabular}

\flushleft*F1 score. Accuracy value not available.
\end{table}

\begin{table}[h!]
\caption{DEAP classification accuracy comparison table}
\label{deap comparison}
  \begin{tabular}{ccc}
    \hline
    \textbf{Method} & \textbf{HV vs. LV (\%)}  &\textbf{HA vs. LA (\%)}\\ \hline
    GaussianNB\cite{koelstra2011deap} & 57.6 & 62.0\\
    RBM with SVM \cite{shu2017emotion} & 60.7 & 64.6\\
    MESAE \cite{yin2017recognition} & 76.2  & 77.2 \\
    DBN \cite{wang2013modeling} & 51.2  & 68.4 \\
    \textbf{HDC early fusion} & 76.7 & 74.2 \\ \hline
  \end{tabular}
\end{table}

One of the key benefits of HDC is the hardware efficiency, which is further improved for large-channeled emotion recognition tasks through the memory optimizations discussed earlier. The results for both valence and arousal accuracy as well as the resulting stored vector count for AMIGOS and DEAP across all memory-optimizing encoding methods are shown in Figure \ref{memopt_perf}. 

For AMIGOS, with 214 channels and 3 modalities, the unoptimized method requires $3$ unique vectors per feature channel \{iM, PFP, NFP\} - a total of $642$ vectors. The `iM vectors constant per modality' scheme is limited by the largest modality which is EEG with $105$ feature channels. This results in a total of $105 + 214\times 2 = 533$ vectors. The `FP constant per feature channel' reduces the total vector set to $105 + 2\times 3 = 111$. The `combinatorial pairs' method uses Equation \ref{combinatorial_equation} and results in a required $31$ vectors. Finally, the `rule 90 generation' stores one FP pair \{PFP, NFP\} for each modality along with the seed vector, a total of $2\times 3 + 1 = 7$. The memory optimizations result in an overall decrease in required vector storage by 98.91\% from $642$ vectors to $7$ while the accuracy actually increased by 1.9\% for arousal and 2.7\% for valence.

For DEAP the overall memory storage is higher due to increased feature channels, $238$, and modalities, $5$. The unoptimized method requires $714$ vectors. The `iM vectors constant per modality' scheme is limited, again, by EEG with $195$ channels totalling $668$ vectors. The `FP constant per feature channel' reduces this to $202$. The `combinatorial pairs' method requires $32$ vectors. Finally, the `rule 90 generation' totals to $11$ vectors. The memory optimizations result in an overall decrease of 98.46\% from $714$ vectors to $11$ while the accuracy actually increased by 0.6\% for arousal and minimally decreased by 1.4\% for valence.

\begin{figure}
  \centering
  \includegraphics[scale=0.22]{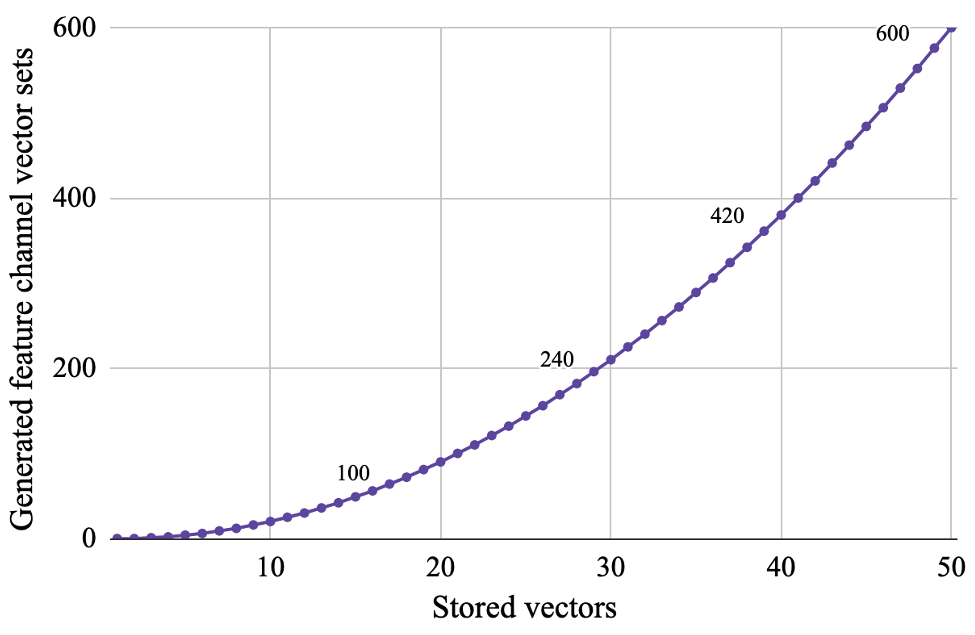}
  \caption{Feature channel vector sets \{iM, PFP, NFP\} generated using combinatorial pair technique vs. stored vectors.}
  \label{vecgenerate}
\end{figure}

Using the combinatorial pair method alone, the relationship between feature channel sets generated and number of stored vectors is shown in Figure \ref{vecgenerate}. 
Linear increases in number of stored vectors will cause result in a quadratically increasing number of available feature channel sets. This plot demonstrates that with $7$ vectors, $9$ feature channel sets are available, but with 50 vectors, $600$ feature channel sets are available.

\begin{figure}
  \centering
  \includegraphics[scale=0.29]{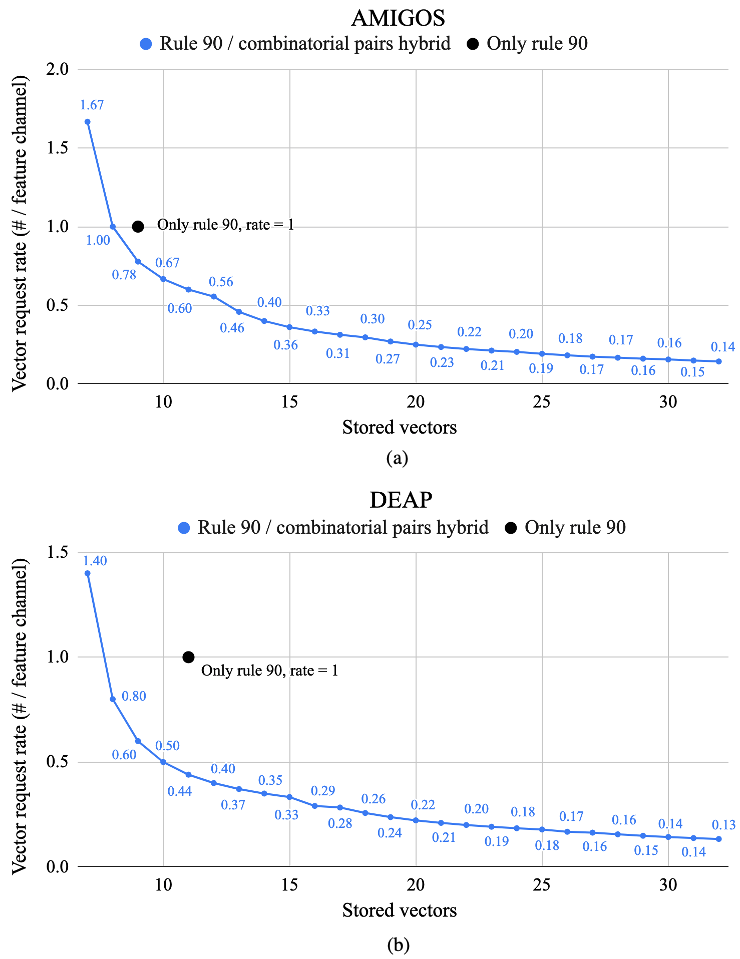}
  \caption{Vector request rate (number of vectors requested per feature channel) vs. memory size (number of vectors) using the hybrid method as compared to only rule 90 for (a) AMIGOS and (b) DEAP}
  \label{vecgen}
\end{figure}

The combinatorial pair method can be used together with rule 90 in a hybrid scheme to provide options for tradeoff between memory and vector requests. In the solely rule 90 version, $7$ vectors are stored for AMIGOS and $11$ vectors for DEAP; a total of 214 and 238 vector requests are made during the encoding process for AMIGOS and DEAP for a single sample - a vector request rate of $1$. However, using the burst generation technique, a small subset of vectors could be generated in one shot using rule 90, stored, and then used for spatially encoding a quadratically larger number of feature channels to reduce the total number of vector requests made. The relationships between vector request rate (total vector requests / number of feature channels) and vector storage for AMIGOS and DEAP are shown in Figure \ref{vecgen}.

The only rule 90 method stores $7$ and $11$ vectors regardless which, if used with the hybrid scheme, could be used to generate $9$ feature channels with every burst instead of only $7$ for AMIGOS, or $25$ feature channels instead of just $11$ for DEAP. This results in a reduction in frequency of vector requests by 22.22\% and 56.00\% for AMIGOS and DEAP, respectively, even while using the same number of stored vectors.

\begin{figure}
  \centering
  \includegraphics[scale=0.31]{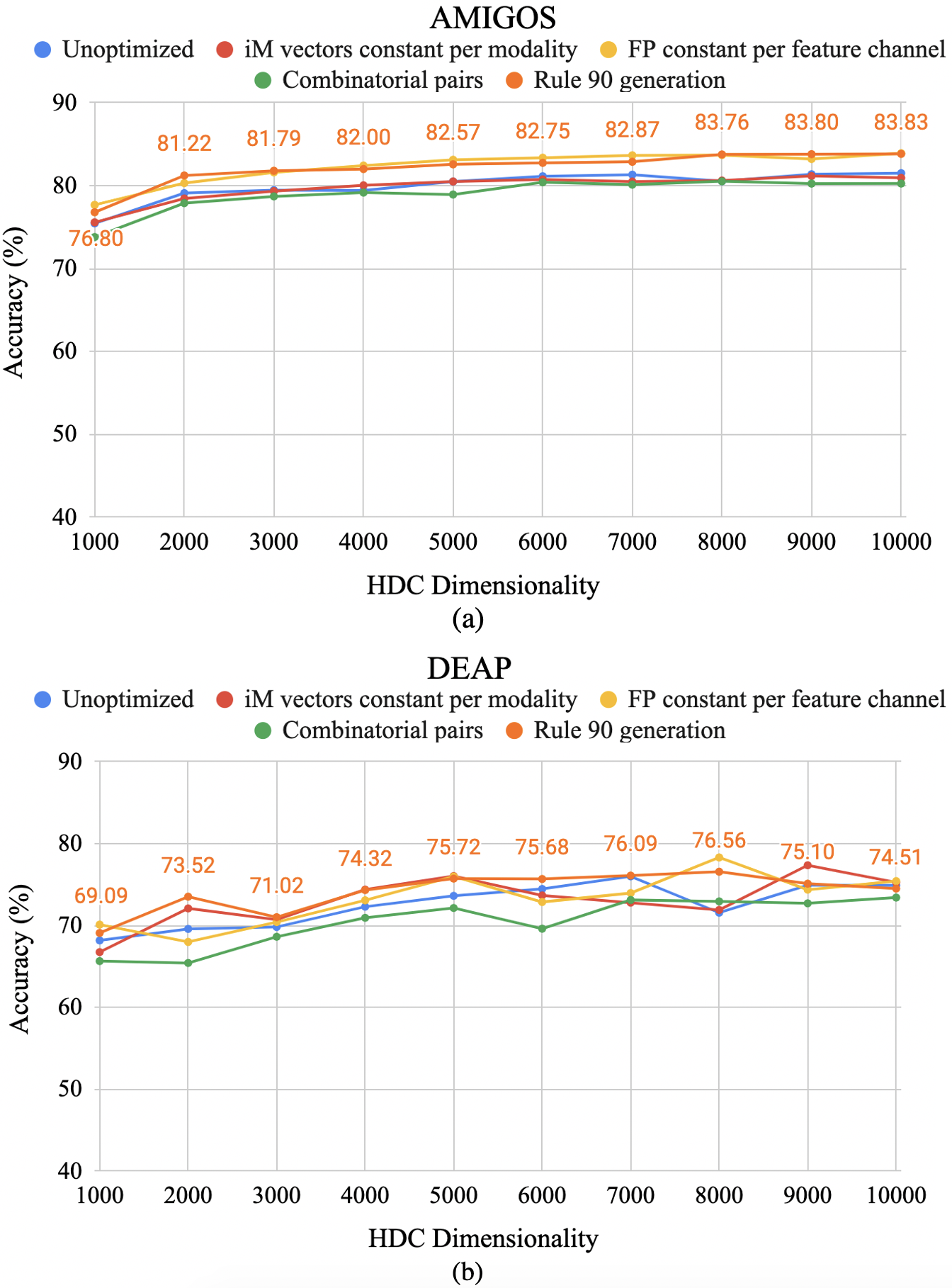}
  \caption{Average accuracy of valence and arousal vs. HDC dimension for the various memory optimizations for (a) AMIGOS and (b) DEAP. The data labels are shown for the most optimized version: rule 90 generation.}
  \label{reduction}
\end{figure}

Finally, the impact of reducing dimension on the overall accuracy performances of the algorithm for emotion recognition tasks are shown for AMIGOS and DEAP in Figure \ref{reduction}. For the AMIGOS dataset, a gradual decrease in accuracy is observed particularly from dimensions of 7000 by which point the average accuracy has dropped by 1\%. Steeper decreases of $\sim$0.6\% and $\sim$4.4\% are seen between dimensions 3000 and 2000 and between 2000 and 1000. Overall, between a dimension of $10,000$ and 1000, there is a decrease in average accuracy of $\sim$7\%.

For the DEAP dataset, there is greater variation in accuracy across the dimensions and between methods, however an overall trend of decreasing accuracy can still be seen, particularly past dimensions of 5000 at which point the accuracy drops below 74.5\% and continues to decrease rapidly including a $\sim$4.4\% drop between 2000 and 1000. Overall, between $10,000$ and 1000 there is a decrease in average accuracy of $\sim$5.4\%.

\section{Discussion}
The first change that was implemented was an overall architecture shift from late to early fusion. The results demonstrate an improvement in performance on the AMIGOS dataset despite moving the fusion point to combine the parallel data streams earlier in the encoding process. The boost in accuracy may come from the fact that different modalities may have different temporal behavior, which may lead to different optimal n-grams. For late fusion, an n-gram of $4$ was used for all modalities without individual tuning. For early fusion, an optimal n-gram of $3$ was selected for the fused modalities temporal behavior, improving the overall performance. The early fusion method requires tuning of only a single temporal encoder and still achieves higher accuracy even with reduction of the overall encoding complexity. This indicates the potential, benefits, and feasibility of early fusion encoding processes in HDC. Information is retained in the high-capacity vectors even with only a single encoding path after the spatial encoder. 

Additionally, compared to other works, as shown in Table \ref{Amigos comparison}, HDC early fusion performed better than GaussianNB, SVM, XGB, ELM and HDC late fusion on AMIGOS. It also performed better, as shown in Table \ref{deap comparison} than GaussianNB, RBM with SVM and DBN and showed similar performance to MESAE on the DEAP dataset. Given its high performance, HDC early fusion appears well-suited for emotion recognition tasks. 

The performance of various memory optimizations were explored and shown in Figure \ref{memopt_perf}. HDC depends on near-orthogonality between different data streams and feature values to ensure that samples from different classes that vary in these ways are encoded into sufficiently orthogonal class vectors. Each optimization reduces the total number of unique vectors that need to be stored in advance, however, there was no decrease in accuracy on AMIGOS between the most unoptimized and most optimized. There was actually an average increase of $\sim$2.3\%; this accuracy change may be attributed to the random element of HDC vector initialization/generation which may result in either beneficial or detrimental random patterns. This is further demonstrated by the DEAP dataset for which the optimizations increased the arousal accuracy by 0.6\%, yet decreased valence accuracy by 1.4\%. With an overall memory reduction of $>$98\%, the optimizations have a significant impact on the hardware requirements while displaying little to no performance degredation for both AMIGOS and DEAP, demonstrating that the techniques generalize across datasets. 

A hybrid, burst generation technique was proposed, in which a small vector set would be used maximally, as shown in Figure \ref{vecgenerate}, and then re-generated. By using this method, the total number of vectors that need to be generated during training or inference of a single sample can be decreased as shown in Figure \ref{vecgen}. Rule 90 alone requires generation of at least one vector per feature channel and doesn't take advantage of the combinatorial pairs available with its existing storage, hence implementation of this hybrid technique decreases the overall required vector generation. The benefit is higher for the DEAP dataset with more modalities due to the prior storage of a larger number of vectors for rule 90.

This technique also allows for scalability while still maintaining memory size; existing vectors pairs can be used to their highest capacity and then the vector bank can be re-generated using rule 90 for the further capacity required by additional channels or modalities. This could be done until the limits of the cellular automata are reached ($>>$10$^3$). The trade-off between the computation for vector generation and additional storage provides options. The optimal performance point based on power or memory constraints can be determined for specific applications/platforms. 

The dimension reduction shown in Figure \ref{reduction} demonstrates the trade-off between accuracy and comprehensive datapath size reduction. An optimal point could be selected that provides the accuracy needs of the system with minimum HDC dimensionality. With a tolerance of $\sim$2\%, the dimension can be reduced by 70\% to hypervectors of 3000 bits for AMIGOS and by 80\% to hypervectors of 2000 bits for DEAP. 

\section{Conclusion}
In conclusion, this work proposed a solution to the many-channeled ($>$200) memory-expensive emotion recognition task in the form of a brain-inspired early fusion hyperdimensional computing architecture alongside several optimization techniques that make emotion recognition feasible for hardware-constrained, low-power wearable applications. The various methods explored were able to achieve significant reduction $>$98\% in required memory and $>$20\% decrease in frequency of vector requests. Finally, the impact of hypervector dimension on emotion recognition accuracy demonstrated $<$2\% performance degradation for datapath reductions of 70-80\%.

Though this work focuses on emotion recognition, all the proposed techniques maintain the properties required for successful hyperdimensional computing and therefore could generalize to other applications, and will be particularly useful for those with many, varied streams of input information. 

To demonstrate the impact of the memory optimizations, the overall area of an ASIC realization of the emotion-classification engine was reduced by 87.3\% for the cellulator automata rule 90 over the unoptimized version in a recent implementation study (unpublished).
Next steps could include implementation of the proposed techniques for other applications with significantly larger numbers of channels and modalities to explore generalizability and scalability. 

\begin{backmatter}
\section*{Abbreviations}
ASIC: application-specific integrated circuit

BVP: blood volume pressure

DBN: deep belief network

ECG: electrocardiogram

EEG: electroencephalogram

ELM: extreme learning machine

EMG: electromyography

EOG: electrooculography

FP: feature projection

GaussianNB: gaussian naive bayes

GSR: galvanic skin response

HCI: human-computer interaction

HD: hyperdimensional

HDC: hyperdimensional computing

HDS: hyperdimensional space

iM: item memory

IoT: internet-of-things

MESAE: multiple-fusion-layer based ensemble classifier of stacked autoencoder

NFP: negative feature projection

PFP: positive feature projection

RBM: restricted boltzmann machine

RFE: recursive feature elimination

SVM: support vector machine

XGBoost: extreme gradient boosting

\section*{Acknowledgements}
The authors would like to thank Abbas Rahimi and Sohum Datta for their interest and support of this work.

\section*{Funding}
This work was supported in part by the CONIX Research Center, one of six centers in JUMP, a Semiconductor Research Corporation (SRC) program sponsored by DARPA, as well as the VIP and HYDDENN programs sponsored by DARPA. This work was supported in part by the National Science Foundation Graduate Research Fellowship Program under Grant No. 1752814. Any opinions, findings, and conclusions or recommendations expressed in this material are those of the author(s) and do not necessarily reflect the views of the National Science Foundation, DARPA or the U.S. government. Support was also received from sponsors of Berkeley Wireless Research Center.

\section*{Availability of data and materials}
Data sharing is not applicable to this article as no new datasets were generated or analysed during the current study.


\section*{Authors' contributions}
A.M., A.N., and R.A. developed and implemented the learning algorithm with optimizations. A.M., D.S., M.A. and H.L. designed the architectures. Y.S.S. and J.M.R. oversaw the project. A.M. and J.M.R. wrote and edited the manuscript.

\section*{Competing interests}
The authors declare that they have no competing interests.



\bibliographystyle{spbasic} 
\bibliography{bmc_article}      

\end{backmatter}
\end{document}